\documentclass{article}

\usepackage{amsfonts,amssymb,amsmath,epsfig}
\usepackage[american]{babel}
\usepackage{graphicx}
\usepackage{subcaption}

\newcommand{\cE}{\mathcal{E}}

\newcommand{\f}{f}

\newcommand{\no}{\noindent}

\newcommand{\al}{\alpha}

\newcommand{\RR}{\mathbb{R}}

\newcommand{\btau}{\boldsymbol{\tau}}

\newcommand{\nup}{\boldsymbol{\nu}_i^{(p)}}
\newcommand{\nus}{\boldsymbol{\nu}_i^{(s)}}
\newcommand{\nuv}{\boldsymbol{\nu}^{(v)}}
\newcommand{\nut}{\boldsymbol{\nu}_i^{(t)}}

\newcommand{\bn}{\boldsymbol{n}}

\newcommand{\bv}{\boldsymbol{v}}
\newcommand{\bw}{\boldsymbol{w}}
\newcommand{\bI}{\boldsymbol{I}}

\def\pb{\, .}
\def\pa{\, \cdotp}
\def\vb{\, ,}
\def\va{\,\raise 2pt\hbox{,}}

\def\bw{{\bf w}}
\def\bx{{\bf x}}
\def\bv{{\bf v}}

\def\bI{{\bf I}}

\def\cB{{\cal B}}
\def\cA{{\cal A}}

\def\cD{{\cal D}}
\def\cC{{\cal C}}
\def\cE{{\cal E}}

\newcommand{\p}{\partial}

\font\dc=cmbxti10 

\begin{document}

\title{Toward a Mathematical Theory of Behavioral-Social Dynamics \\
       for  Pedestrian Crowds}

\author{Nicola Bellomo\\
\small Department of Mathematics, Faculty  Sciences, King Abdulaziz University\\[-0.8ex]
\small Jeddah, Saudi Arabia \\ 
\small Address: Politecnico di Torino, 10129 Torino, Italy \\       
\small \texttt{nicola.bellomo@polito.it}\\      
\and
Livio Gibelli\\
\small Department of Mathematical Sciences, Politecnico di Torino \\[-0.8ex]
\small Corso Duca degli Abruzzi 24, 10129 Torino, Italy \\
\small \texttt{livio.gibelli@polito.it}
}


\maketitle

\begin{abstract}
This paper presents a new approach to  behavioral-social dynamics of pedestrian crowds by suitable development of methods of the kinetic theory.
It is shown how heterogeneous individual behaviors can modify the collective dynamics, as well as how local unusual behaviors can propagate in the 
crowd. The main feature of this approach is a detailed analysis of the interactions between dynamics and social behaviors. \\

\end{abstract}

\section{Plan of the Paper}

This paper deals with the  modeling of {\dc behavioral-social crowd dynamics}, where this term is used to indicate that laws of classical mechanics can be substantially modified by individual behaviors and strategies developed by living entities. Namely, the individuals in the crowd, are viewed as self-propelled micro-systems, who can modify their collective behaviors according to their walking strategies. These micro-scale subsystems are featured by an heterogeneously distributed ability to express their walking strategy related to the interactions with the other entities.

The paper aims at presenting a unified approach, to modeling and simulation, based on the methods of the  kinetic theory and stochastic games \cite{[BKS13]}  which, according to the authors' bias, offer an appropriate framework to capture the greatest part of the complexity features of the systems under consideration.  Classical methods of the kinetic theory for molecular fluids cannot be straightforwardly applied. Indeed,  conservation of number of particles can be claimed  in the case of self propelled particles, but not conservation of momentum and energy. The modeling involves several technical difficulties. 
Understanding these difficulties is a necessary preliminary step to develop a successful approach.

The interested reader can take advantage of some review papers, which enlighten different aspects of the modeling of crowd dynamics. Restricting our attention to the most recent literature,   the survey paper  \cite{[BPT12]}  presents and critically analyzes a variety of crowd models at all representation scales, from the microscopic (individual) to macroscopic (hydrodynamical). A mathematical approach to the modeling of swarm dynamics, which has some intersection with crowd dynamics, is proposed in \cite{[BS12]}.

Paper  \cite{[BBK13]} is proposed  for crowd dynamics in unbounded domain with homogeneous distribution of walking ability of pedestrians. The present paper aims at including more general features of  behavioral-social  dynamics, as well as interactions with walls. Moreover, the approach of  \cite{[BBK13]} uses discrete velocity variables, while a continuous variable is used in this present paper. This choice allows to overcome  the uncertainty
problem induced by the choice, up to now heuristic, of the number of velocity module and directions.
The contents of the paper are as follows:

Section 2 defines an assessment of the basic  principles of behavioral-social dynamics of a large systems of interacting self-propelled particles with focus on crowd dynamics. The analysis is related to the complexity  features of these specific systems referring to a  scaling and representation, which anticipates the  mathematical kinetic theory for active particles \cite{[BKS13]}. The microscopic state of the latter, which constitute  the micro-scale system, whose micro-state includes not only mechanical variables, typically position and velocity, but also an additional variable, called activity, suitable to model their strategy.

Section 3 develops the concept of  {\dc behavioral--social dynamics} as a new science of mechanics, where interactions depend on the behaviors of the entities of the crowd, which is viewed as a living system. A general structure is used toward the derivation of specific models. The approach consists in representing the state of the system by a probability distribution over the micro-scale state of pedestrians and in deriving a general framework to describe  the time and space dynamics of such distribution by a balance of particles in the elementary volume of the space of the micro-states. Interactions, which are non local and nonlinearly additive, are modeled by theoretical tools of stochastic game theory.

Section 4 takes advantage of the framework presented in the preceding section to derive models which include also the case of individuals which change the rules of their participation to the dynamics.  The whole section is in three parts. First the dynamics is modeled in unbounded domains. Then it is shown how the decision process by which individuals select their trajectories include the actions to avoid walls and obstacles. Finally, it is shown   their social behavior evolves in time and space focusing on  panic conditions \cite{[HFV00],[HJA07]}.

Section 5 presents some simulations with to test the predictive ability of the model. Particles methods (Monte Carlo type) are used as this approach appears to be well consistent with the general mathematical structure proposed in Section 3. Two types of simulations are developed, the first one to show how the individuals moving in opposite directions interact, while the second one aims at understanding how panic conditions can affect the evacuation dynamics. Paper \cite{[HJ09]}  is an important reference to define specific objective of  simulations.

Section 6 presents a critical analysis and indicates how the approach can be further developed to include additional aspects of the dynamics such as modeling interactions which induce large deviations in the social behavior of the crowd. This section also outlines some research perspectives focusing
on a systems approach to crowd dynamics in complex environments.


\section{Behavioral-Social  Dynamics of Self-Propelled Particles}

This section provides an heuristic description of behavioral dynamics of pedestrian crowds to be transferred into appropriate mathematical structures. of the mesoscopic (kinetic) theories, where the micro-state of pedestrians is given by position and velocity, while the system is represented a  probability distribution function over such state. Mathematical models describe the time evolution of this function in the phase space by means of integro-differential equations. The {\dc micro-scale description}, where pedestrians are identified singularly, is delivered by position, velocity, and additional variables suitable to retain features of pedestrians viewed as living entities.

Let us now understand   what is a crowd and which are the most important features of a crowd to be taken into account in the modeling approach. First a definition of the {\dc crowd} needs to be given. The following one (due to  Helbing and Johanson \cite{[HJ09]}) appears to be particularly  well focused:
\begin{quote}
\textit{Agglomeration of many people in the same area at the same time. The density of the crowd is assumed to be high enough to cause continuous interactions with or reaction to other individuals.}
\end{quote}

According to the authors' bias, only three important specific features of the crowd  are selected among various ones. Subsequently, a critical analysis of the existing literature will focus on topics to be considered open, and to be treated in the next sections.

\vskip.1cm $\bullet$  \noindent  {\dc Strategy:} Pedestrians have the ability to express walking strategies based on interactions with other individuals and with the surrounding environment. The latter includes  vocal and visual signaling  which addresses them toward optimal and safe ways, including evacuation paths.
 \textit{The modeling of pedestrians' strategy should include several features, for instance trend toward the exit or a meeting point, following or avoiding streams and clusters, avoiding overcrowding in the proximity of walls, clustering of individuals with similar activity, avoiding individuals with different activity, and possibly others}.

\vskip.1cm  $\bullet$  \noindent   {\dc Heterogeneity:} The behavior of pedestrians is heterogeneously distributed due both to different  psychologic attitudes and mobility abilities  from individuals with handicap to high level walking ability. In addition, a crowd might need to to be split into different groups related to different strategies, e.g. reaching different objectives, or even an internal hierarchy, which induces different interaction rules.
\textit{Heterogeneity can refer to social behaviors, for instance aggressiveness in a crowd where two groups contrast each other or panicking behaviors.  Indeed, crowds can lose, in panic conditions, optimal strategies. This feature can be important in extreme situations when sudden dangers can induce the said conditions}.

\vskip.1cm   $\bullet$  \noindent   {\dc Interactions:}  Interactions involve both  mechanical and social-behavioral features, and are nonlocal as individuals communicate and develop a visual activity  at a distance; these are nonlinearly additive as the strategy developed by a pedestrian is a nonlinear combination of different stimuli, while mechanics induces social exchanges and these modify the walking dynamics.
 \textit{Interactions with the external environment where the pedestrians move, namely different geometrical and environmental contexts, say corridors, rooms,  stairs, sudden changes of directions, luminosity conditions, and many others, can have an important effect on the dynamics as the interaction rules, which also depend on the quality of the environment}.

\vskip.1cm

Although only three general features have been selected, one can rapidly verify that the existing literature does not yet provides an exhaustive answer to all of them. In fact, recent papers, e.g. \cite{[ACK14],[BBK13]}, only treat some of the aforesaid topics, while  the interplay between mechanics and social behaviors deserves further attention. 
The meso-scale approach, introduced in \cite{[BB14]} and further developed in \cite{[ACK14]}, shows how pedestrians, viewed as heterogeneous entities chose, by a decision process modeled by tools of game theory, between different trends  from avoiding walls to reach prescribed exits. It is an approach quite different from that developed at the microscopic scale, for instance by the {\dc social force} model \cite{[HEL01]}, which is  based on the assumption that pedestrians are subject to an acceleration to be related to social interactions within the crowd. However, models at the microscopic scale can contribute not only to understand how individual behaviors can be described by equations, but also to model interactions. A useful example is given by the paper by Faure and Maury \cite{[FM15]}, which provides a detailed analysis of the granular dynamics based on different aspects of attraction and repulsion between pedestrians.

 The  next sections present a model, which should at least partially, include the predictive ability of the aforesaid features. The derivation will be followed by a critical analysis the said ability also referred to validation issues. This critical analysis can take advantage of the literature in the field, which already offers valuable contributions on theoretical models and  interpretation of empirical data, among others, \cite{[DH03],[MHG09],[MT11],[SS11],[SSKB06]}.

 Finally, let us mention the strategic motivation to study anomalous behaviors and, in particular, the onset and propagation of panic conditions. This study is motivated by the related safety problems, see \cite{[HFV00],[HJA07]} for a deeper understanding of a psycho-mechanical study of this extreme phenomenon and \cite{[ACK14]} for a detailed analysis of evacuation times and the impact that panic conditions can have on it. Indeed, the study of this feature is an important issue of this present paper.


\section{On the Kinetic Theory Approach to Behavioral Dynamics}

Let us consider a large heterogeneous system of pedestrians moving in two dimensional  domains. The mathematical approach to modeling
cannot relay, as observed in~\cite{[BKS13]}, on the deterministic causality principles typical of classical mechanics. 
In fact, pedestrians develop a their own dynamics based on an individual interpretation of that of the other individuals. 
Namely, they have a strategy, which is heterogeneously distributed and which depends on several factors to be included in the modeling approach.
These reasonings lead to introduce the concept of  {\dc behavioral-social  dynamics} as a new science of mechanics, where interactions depend on the 
behaviors of the component of the crowd viewed as a living system. This section proposes a general structure to be used toward the derivation of 
specific models.\\
Bearing this in mind, let us anticipate some terminology and some preliminary ideas of the approach that will be developed hereinafter.

\begin{itemize}

\item The modeling approach proposed in this paper is based on suitable developments of the so-called kinetic theory for active particles, 
      which applies to large systems of interacting entities~\cite{[BKS13]}.  Hence the meso-scale representation is chosen.

\item Pedestrians, namely  the  {\dc micro-system}, are viewed as {\dc active particles}, that have the ability of expressing a their own strategy, 
      called {\dc activity}. This ability can differ for different groups in the same crowd, being understood that the activity is heterogeneously 
      distributed.

\item The overall state of the system is described by a probability distribution over the variable deemed to define the physical state of the said 
      entities, while interactions are modeled by theoretical tools of game theory~\cite{[NOW06]}. 
      A general overview of this approach is presented in the survey paper~\cite{[BKS13]}, while applications to model crowd dynamics and social 
      systems are proposed in~\cite{[BBK13],[BB14]}.

\item The overall system is subdivided into groups of pedestrians who share common features. The partition follows the hallmarks of a systems theory of social systems introduced in \cite{[ABE08]} and followed in \cite{[BHT13],[DL14]}.

\item Active particles can communicate and develop a social dynamics, as they learn from interactions and accordingly modify both strategy and  dynamical rules followed in the movement. The output is a collective behavior which can be observed in the whole.

\end{itemize}

Modeling  is developed in three steps, each of them presented in the next subsections, namely representation of the system, modeling of micro-scale interactions, and derivation of a mathematical structure consistent with the complexity paradigms of behavioral dynamics.  This last subsection also proposes  a concise critical analysis.

\subsection{Representation}
Let us now consider a crowd which can be subdivided into different groups distinguished either by  different walking objectives and/or abilities. 
Pedestrians move in the whole plane $\RR^2$,  but at an initial time  are located in a domain $\Sigma$. 
Each group, called {\dc functional subsystem},  has a different walking purpose, for instance either following a certain direction or reaching a 
meeting point.

The {\dc  microscopic state} of pedestrians, viewed as {\dc active particles}, is defined by position $\bx$,  velocity $\bv$,  and activity $u$. 
Dynamics in two space dimensions is considered, while polar coordinates are used for the velocity variable, namely $\bv = \{v, \theta\}$, 
where $v$ is the velocity modulus and $\theta$ denotes the velocity direction. 

Moreover, we assume that each group, called {\dc functional subsystem}, can be featured by a different group's strategy, ability to express it is  modeled by a variable $u\in[0,1]$, called {\dc activity}, such that $u=0$ denotes the worst walking ability, while $u=1$ the best one. Moreover, the quality of the environment is also taken into account by a parameter $\al \in [0,1]$, where the lower value $\al =0$ denotes the worse environmental conditions that prevents the movement, while  $\al=1$ the best one.

Dimensionless, or normalized, quantities are used  by referring the components $x$ and $y$  to the length $\ell$ that corresponds to the largest dimension of $\Sigma$, while the velocity modulus is divided by  the maximum admissible velocity $V_\ell$, which can be reached by a fast pedestrian in free flow conditions; $t$ is the dimensionless  time variable obtained referring the  real time  to a suitable critical time $T_c$ identified by the ratio between $\ell$ and $V_\ell$.

The {\dc mesoscopic (kinetic) representation} of the overall system  is delivered  by the statistical distribution at time $t$, over the micro-scale state:
\begin{equation}
	f_i=f_i(t,\,\bx,\,v, \theta, u), \quad \bx\in\Sigma,\quad v\in [0,1], \quad \theta \in [-1,1) \quad u\in [0,1],
\end{equation}
for each functional subsystem labeled by  $i = 1, \ldots, n$, and denoted by the acronym  $i$-fs.

\vskip.1cm \no \textbf{Remark 3.1.} \textit{Functional subsystems can be identified by different features, for instance trend to different exits or directions, different roles in the crowd such as hierarchical behaviors. Examples will be given in the applications proposed in the following.}
\vskip.1cm

If $f_i$ is locally integrable then $f_i(t,\,\bx,\,v, \theta, u)\,v\, dv\, d\theta \,du$ is the (expected) infinitesimal number of pedestrians who, at time $t$, have a micro-state comprised in the elementary volume of the space of the micro-states of each  $i$-fs. The following compact expression $\bw = \{v, \theta, u\}$, with $\bw \in D_\bw$, is occasionally used in the following so that $f_i=f_i(t,\,\bx,\,v, \theta, u)= f_i(t, \bx, \bw)$, while the $f_i$ are divided  by  $n_M$, which which defines  the maximal full packing density of pedestrians. Although their state is represented by a point, pedestrians are supposed to have a finite dimension.

{\dc Macroscopic observable quantities} can be obtained, under suitable integrability assumptions, by weighted  moments of the distribution functions:
\begin{equation}
\rho_i (t, \bx)  =  \int_0^1 \int_{-1}^1  \int_0^1  f_i(t,\,\bx,\,v, \theta, u)\,v dv\, d\theta\,du,
\end{equation}
 while analogous calculations lead to the flux
\begin{equation}
{\bf q}_i (t, \bx)  =  \int_0^1 \int_{-1}^1 \int_0^1   \bv\, f_i(t,\,\bx,\,v, \theta, u)\,v dv\, d\theta\,du.
\end{equation}
Global quantities are obtained by summing over the index labeling the functional subsystems. For instance,
mean velocity and velocity variance are computed as follows:
\begin{equation}
\boldsymbol{\xi} (t,\bx) = \frac{1}{n(t, \bx)} \, \sum_{i=1}^n\, \xi_i(t,\bx) = \frac{1}{n} \, \sum_{i=1}^n\, \frac{{\bf q}_i (t, \bx)}{\rho_i (t, \bx)}
\end{equation}
and
\begin{equation}
\sigma (t, \bx)  =\frac{1}{n}\,  \sum_{i=1}^n\,  \int_0^1 \int_{-1}^1 \int_0^1  (\bv - \xi (t,\bx))^2 \, f_i(t,\,\bx,\,v, \theta, u)\,v dv\, d\theta\,du.
\end{equation}

An additional quantity to be taken into account in the modeling of interactions is the \textit{perceived density} $\rho_\theta^a$ along the 
direction $\theta$. According to~\cite{[BB14]}, this quantity is defined as follows:
\begin{equation}\label{perceived}
\rho^a_\theta  = \rho^a_\theta [\rho]= \rho + \frac{\partial_\theta \rho}{\sqrt{1 + (\partial_\theta \rho)^2}}\,\big[(1- \rho)\, H(\partial_\theta \rho) + \rho \, H(- \partial_\theta \rho)\big]\va
\end{equation}
where $\partial_\theta$ denotes the derivative along the direction $\theta$, while  $H(\cdot)$ is the Heaviside function,
$H(\cdot \geq 0) = 1$, and $H(\cdot < 0) = 0$. Therefore, positive gradients increase the perceived density up to the limit $\rho = 1$, while negative gradients decrease it down to the limit $\rho = 0$ in a way that
$$
\partial_\theta \rho \to \infty \Rightarrow \rho^a \to 1\vb \quad \partial_\theta  \rho = 0 \Rightarrow \rho^a = \rho \vb \quad \partial_\theta  \rho \to - \infty \Rightarrow \rho^a \to 0\pb
$$

\subsection{Hallmarks toward modeling interactions}

As already mentioned, interactions may depend not only on the micro-state of the interacting particles (pedestrians), but 
also on their distribution functions. When only the first type of interaction appears, one talks about {\dc linear interactions}, 
otherwise when also second type occur, the concept of {\dc nonlinear interactions} needs to be used.
While linearity involves only independent variables, i.e., micro-state variables, nonlinearity involves the dependent variables, i.e.,
the distribution function and/or its moments. 
In the following, round and square parenthesis distinguish, respectively, the 
former and latter interactions.

The modeling of interactions corresponds to a decision process for each particle related to the micro-state, and distribution function, of the particle and of all those in its  interaction domain. For each functional subsystem, three types of particles are involved in the process at each time $t$: The {\dc test particle} with micro-state $\bw$ and distribution function $f_i(t,\bx, \bw)$; the {\dc field particle}, in $\bx^*$, with micro-state $\bw_*$ and distribution function $f_i(t, \bx, \bw^*)$; and {\dc candidate particle}, in $\bx$, with micro-state $\bw^*$ and distribution function $f_i(t, \bx, \bw_*)$. The candidate particle can acquire, in probability, the micro-state of the test particle after interaction with the field particles, while the  test particle loses its state in the  interaction with the field particles. The test particle is representative, for each functional subsystem, of the whole system of particles.

Interactions of test and candidate particles with field particles, can be modeled by the following quantities: {\dc interaction domain} $\Omega$, {\dc interaction rate} $\eta$, and {\dc transition probability density} $\cA$. These quantities can depend, as already mentioned, on the micro-state and distribution function of the interacting particles, as well as on the quality of the environment. Moreover, they refer to interactions involving all functional subsystems. The expression of these terms is reported in the following, where the term $i$-particle is used to denote a pedestrian belonging to the $i$-th functional subsystem.

\begin{itemize}

\item {\dc Interaction domain:} Active particles have an interaction domain $\Omega$ to be related to their visibility domain which   can be defined as an  arc of circle, with radius $R$ and $R_V$, symmetric with respect to the velocity direction being defined by the visibility angles $\Theta$ and $- \Theta$. For practical applications and according to the normalization, one can assume $\Omega \cong [-\frac{1}{3},\frac{1}{3}]$ corresponding to a visibility of $120$ degrees.

\vskip.2cm \item {\dc Interaction rate:} This term models the interaction rate between a candidate h--particle (or test) in $\bx_*$  (or in $\bx$) and a field k--particle in $\bx^* \in \Omega$.   The following notation, referred to candidate and field particles, can be used $\eta_{hk}[f](\bx_*,\bx^*,\bw_*,\bw^*; \alpha)$. Analogous notation is used for a test particle in $\bx$ and a field particle in $\bx^*$.

\vskip.2cm \item {\dc Transition probability density:}  Interactions can be modeled by the probability density $\cA_{hk}^i[f](\bw_* \to \bw|\bw_*, \bw^*;\alpha)$, which
models the probability that a candidate h-particle with state $\bw_*$  in $\bx_*$ falls into the state  $\bw$ due to the interaction with  a field k--particle in $\bx^* \in \Omega$ with state  $\bw^*$.

\end{itemize}

\vskip.1cm \no \textbf{Remark 3.2.} \textit{When the system does not allow transitions from across functional subsystems the following notation is
used $\cA_{ik}^i =: \cA_{ik}$, while when interactions do not depend on the other subsystems one has  $\cA_{ii}^i =: \cA_{ii}$.}

\vskip.1cm \no \textbf{Remark 3.3.} \textit{In general, both  $\eta_{hk}$ and $\cA_{hk}^i$ can depend  on the micro-scale states of the interacting particles and on the density of the field particles in the domain $\Omega$. Moreover interactions depend on the quality of the environment which can be modeled by the parameter $\alpha \in [0,1]$, where $\alpha = 0$ correspond the worse conditions preventing the dynamics, while $\alpha = 1$ to the best ones, which allow a rapid dynamics. This type of nonlinearity will be discussed later.}
\vskip.1cm

\subsection{A mathematical structure}

The mathematical framework consists in an integro-differential equation suitable to describe the time dynamics of the distribution functions $f_i$, which can be obtained by a balance of particles in the elementary volume of the space of the micro-states. This conservation law writes:
\begin{eqnarray*}
&&\hbox{\bf Variation rate of the number of active particles}\\
&& \hskip1truecm = \hbox{\bf Inlet flux rate - Outlet flux rate},
\end{eqnarray*}
where the inlet and outlet fluxes are caused by interactions. This equality corresponds to the following structure:
\begin{equation}
\label{eq:kin}
\left(\p_t  +  \bv \cdot \p_\bx \right)\, f_i(t, \bx, \bw) = J_i[\f](t, \bx, \bw; \alpha) = \big(G_i -  L_i \big)[f](t, \bx, \bw; \alpha),
  \end{equation}
where the various terms $J_i$  can be formally expressed, consistently with the definition of $\eta$, and $\cA$. The formal result, in the case of individuals who do not move from one functional subsystem to the other,  is as follows:
\begin{eqnarray}
G_i &=& \sum_{k=1}^n \,  \int_{\Omega \times D_\bw^2} \eta_{ik}[\f](\bw_*,\bw^*; \alpha)
 \, \cA_{ik}[\f] (\bw_* \to \bw |\bw_*, \bw^*, u_*;  \alpha) \\
\label{eq:gain}
   &\times&   f_i(t, \bx, \bw_*) f_k(t, \bx^*, \bw^*)\,d\bw_*\,d\bw^* \, d\bx^*,
 \end{eqnarray}
 and
\begin{equation}
\label{eq:loss}
L_i =\sum_{k=1}^n f_i(t, \bx, \bv) \int_{\Omega\times D_\bw } \eta_{ik}[\f](\bw_*,\bw^*; \alpha)\, f_k(t, \bx^*, \bw^*)\, d\bw^* \, d\bx^*.
\end{equation}

When the aforesaid crossing over functional subsystems, induced by  change of behaviors, are allowed, the interaction term of the structure modifies as follows:
\begin{eqnarray}
G_i &=&\sum_{h=1}^n \, \sum_{k=1}^n \,  \int_{\Omega \times D_\bw^2} \eta_{hk}[\f](\bw_*,\bw^*)
\, \cA_{hk}^i[\f] (\bw_* \to \bw |\bw_*, \bw^*, u_*) \nonumber \\
\label{eq:gain2}
 &\times&   f_h(t, \bx, \bw_*) f_k(t, \bx^*, \bw^*)\,d\bw_*\,d\bw^* \, d\bx^*,
 \end{eqnarray}
and
\begin{equation}
L_i =\sum_{k=1}^n f_i(t, \bx, \bv) \int_{\Omega\times D_\bw } \eta_{ik}[\f](\bw_*,\bw^*)\, f_k(t, \bx^*, \bw^*)\, d\bw^* \, d\bx^*.
\end{equation}

Finally, it can be briefly shown that this structure is consistent with the paradigms presented in Section 2.  The ability of pedestrians to express walking strategies based on interactions with other individuals is modeled by the transition probability density, while  the heterogeneous distribution of the said  strategy (behavior) corresponding both to different  psychologic attitudes and mobility abilities is taken into account by the use of a probability distribution over the activity variable.  Interactions have been assumed to be nonlocal and nonlinearly additive as the strategy developed by a pedestrian is a nonlinear combination of different stimuli generated by interactions with other pedestrians and with the external environment.


\section{Mathematical Models}

The hallmarks to derive models by the kinetic theory for active particles are reported in \cite{[BKS13]}, where the hallmarks proposed in 
this present paper can be followed. In detail, first a general structure suitable to capture the main complexity features of living systems 
is derived and subsequently such a structure is implemented by models of interactions at the micro-scale. 
This approach aims at overcoming the lack of first principles that govern the living matter. 
Specific models are derived in this section within the framework proposed in the preceding one. 
The approach first refers to Eqs.~(3.7),(3.8),(3.9) focusing to well defined case studies, where active particles do not move from one subsystem 
to the other; subsequently, it is shown how models can be generalized to include also the aforesaid transition as by  Eqs.~(3.7),(3.10),(3.11). 
The next two subsections are devoted to this objective, while the third one presents a critical analysis focused on the modeling of panic conditions, 
and more in general anomalous behaviors.

\subsection{Models in unbounded domains}
Let us now consider the modeling of crowd dynamics when pedestrians have a well defined strategic objective and do not cross from one functional
subsystem to the other. In this specific case, the activity variable corresponds to the walking ability and the derivation of models needs defining
the terms $\eta_{ik}$ and $\cA_{ik}$, which describe interactions at the micro-scale. First some phenomenological assumptions on the qualitative
micro-scale dynamics are made and subsequently these are transferred into models to be implemented into the mathematical structure.
In detail:
\begin{enumerate}
\item The encounter rate corresponds to the frequency by which each pedestrian develops contacts with the other ones in the visibility zone $\Omega$.

\vskip.2cm \item Interactions modify the dynamics of pedestrians who first might change the direction of movements, and subsequently adapt their velocity to the local density conditions.

\vskip.2cm \item Three types of stimuli contribute to the modification of walking direction:
                (i) desire to reach a well defined target, namely a direction or a meeting point;
                (ii) attraction toward the mean stream;
                (iii) attempt to avoid overcrowded areas.

\vskip.2cm \item Pedestrians moving from one direction to the other, adapt their velocity to the local new density conditions,
                 namely they decrease speed for increasing density and increase it for decreasing density.

\vskip.2cm \item The dynamics is more rapid in amply and high quality areas. Moreover rapidity is heterogeneously distributed and increases 
                 for high values of the activity variable.

\vskip.2cm \item Interactions also modify the activity variable according to a social dynamics based on attraction and/or repulsion of social behaviors.
\end{enumerate}
Bearing all above in mind, let us transfer this qualitative description into mathematical models of a micro-scale dynamics.
According to these phenomenological assumptions, the gain and loss terms in the kinetic equation, Eqs.~\eqref{eq:gain}
and \eqref{eq:loss} respectively, simplify to:
\begin{eqnarray}
\label{eq:gain_simplified}
G_i & = & \int_{\Omega \times D_\bw} \eta_{i}[\f](\bw_*; \alpha) \;
      \cA_{i}[f] (\bw_* \to \bw |\bw_*, u_*;  \alpha) f_i(t, \bx, \bw_*)\,d\bw_*, \\
\label{eq:loss_simplified}
L_i & = & \int_{\Omega \times D_\bw} \eta_{i}[\f](\bw_*; \alpha)\,d\bw_* \; f_i(t, \bx, \bw).
\end{eqnarray}

\vskip.1cm \noindent \textbf{Modeling the encounter rate, $\eta_{i}$.}
In general, the encounter rate can depend on the micro-state and the meso-state of the interacting individuals viewed as active particles.
A rule generally adopted is that the frequency of interactions depends  on the number of particles in the interaction domain. However, the
conjecture that individuals consider only a fixed number of individuals to develop the decision process on their walking strategy was posed
in~\cite{[BCC08]}. See also the general formalization of~\cite{[BS12]} and the computational hints in~\cite{[AP12]}.
Therefore, if the visibility allows to capture this critical density, a reasonable assumption is simply $\eta \cong \eta_0$.
However, lack of visibility or presence of obstacles can reduce this encounter rate.

\vskip.1cm \noindent \textbf{Modeling the transition probability density, $\cA_{i}$.}
Let us consider a  candidate pedestrian located in $\bx$ with velocity direction $\theta_*$ and modulus $v_*$, and let us define the following
unit vectors:
\begin{itemize}
 \item $\nut(\bx)$ directed to a prescribed meeting point or to a walking direction;
\vskip.2cm \item $\nus[\f](\bx)$ along the local stream (we assume the mean velocity but other choices can be made, i.e., mean flux);
\vskip.2cm  \item $\nuv[\f](\bx)$ directed along the direction of lowest density gradient;
\end{itemize}
where both $\nus$ and  $\nuv$ are computed within the visibility zone.

\noindent
Pedestrians in addition to the trend to  $\nut$, which depends on the position only, develop a decision process between the attraction
by the stream, namely  $\nus$, and the search of a low density path  $\nuv$, where both unit vectors depend on $\bx$ and on $\f$ 
through velocity and density, respectively.
Paper~\cite{[ACK14]} suggests that a preferred direction can be  heuristically chosen. Accepting this hint, the following model of
{\dc preferred direction} is proposed:
\begin{equation}\label{nu}
\nup =  \frac{(1-\rho) \nut + \rho \left[ \beta \nus + (1-\beta) \nuv \right]}
               {\left\| (1-\rho) \nut + \rho \left[ \beta \nus + (1-\beta) \nuv \right] \right\|},
\end{equation}
where $\beta$ is a parameter which models the sensitivity to the stream with respect to the search of vacuum.

\vskip.1cm \no \textbf{Remark 4.1.} \textit{If the crowd has not a well defined trend to a prescribed direction or to a meeting point, 
                                            then the dynamics is simply ruled by the remaining two trends and the preferred direction is:
\begin{equation}\label{nu_bis}
\nup =  \frac{\beta \nus + (1-\beta) \nuv}
               {\left\| \beta \nus + (1-\beta) \nuv \right|}\pa
\end{equation}}

\vskip.1cm \no \textbf{Remark 4.3.} \textit{A correspondence can be easily identified, for practical simulations, between the preferred direction 
                                            $\nup$ and the preferred angle of motion, $\theta_i^{(p)}$, i.e., 
                                            $\nup=(\cos{\theta_{i}^{(p)}},\sin{\theta_{i}^{(p)}})$.}

\vskip.1cm
All tools have now been defined to model the decision process leading to the transition probability density, $\cA_{i}$, which is defined as follows
\begin{equation}
\label{eq:transition_probability}
\cA_i[\f](t, \bx, \bw; \alpha) =
\cD [\rho] ( u_* \to u )
\cC [\rho] ( v_* \to v )
\cB_i [\rho,\mathbf{q}_i] (\theta_* \to \theta).
\end{equation}

\begin{figure}[h]
   \begin{center}
     \epsfig{file=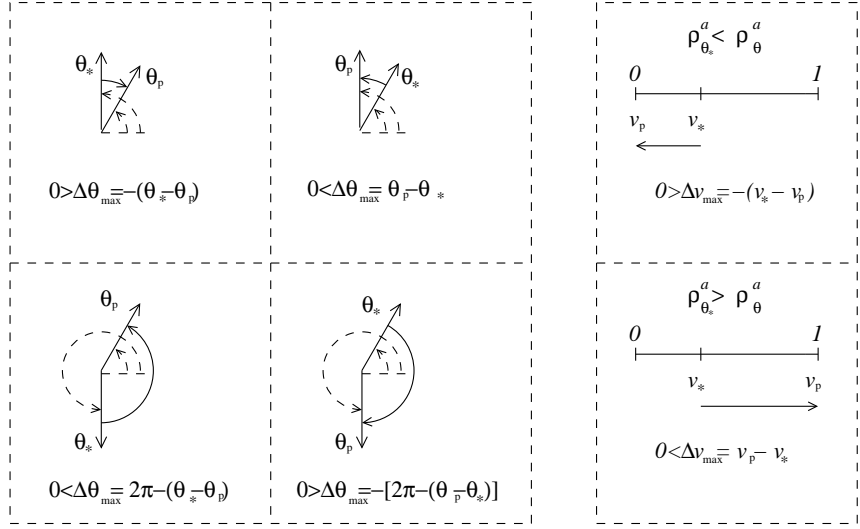,height=7cm}
   \end{center}
\caption{Explicit expressions of $\Delta \theta_{max}$ and $\Delta v_{max}$.}
\label{fig:p_transizione}
\end{figure}

More in detail:
\begin{enumerate}
\item The candidate pedestrian in $\bx$ with velocity direction $\theta_*$ changes in probability  $\theta_*$ into $\theta$ depending on the
      direction $\theta_i^{(p)}$, on the local density, activity, and quality of the environment;
      
\vskip.2cm \item After having changed velocity direction  the velocity is modified by increasing (decreasing) its modulus depending on whether 
                 the perceived density is lower (higher) than the previous one;
      
\vskip.2cm \item Finally, the candidate pedestrian modifies the activity by social communications.
\end{enumerate}
According to the aforesaid phenomenological description, the transition probability density for angles is assumed to vary linearly from
$\theta_*$ to $\theta_i^{(p)}$, i.e.,
\begin{equation}
\label{eq:B}
\cB_i[\rho,\boldsymbol{q}_i]\left(\theta_* \to \theta=\left(\theta_* + \Delta\theta \right)\;\mbox{mod}\;2\pi\right) =
2 \frac{\phi_\theta-\psi_{i,\theta}}{\Delta \theta_{i,max}^2} |\Delta\theta| +
\frac{1+\psi_{i,\theta}-\phi_\theta}{|\Delta \theta_{i,max}|},
\end{equation}
where $\Delta\theta \in \left[0,\Delta\theta_{i,max}\right]$ and the maximum possible variation of velocity direction is given by
\begin{equation}
\Delta\theta_{i,max}=(\theta_i^{(p)}-\theta_*)-2\pi\;\mbox{sign}{(\theta_i^{(p)}-\theta_*)}\;H(|\theta_i^{(p)}-\theta_*|-\pi)
\end{equation}
where $H(\cdot)$ is the Heaviside function.
In Eq.~\eqref{eq:B}, $\phi_\theta$ and $\psi_\theta$ give the negative and positive contributions to the trend to $\theta_i^{(p)}$,
respectively,
\begin{equation}
 \phi_\theta  = \rho, \qquad \hbox{and} \qquad  \psi_{i,\theta} = \alpha u_* \frac{|\Delta\theta_{i,max}|}{\pi}
\end{equation}
Likewise, the transition probability density for the velocity modulus is given by
\begin{equation}
\label{eq:C}
\cC[\rho]\left(v_* \to v=v_* + \Delta v \right) =
2 \frac{\phi_v -\psi_v}{\Delta v_{max}^2} |\Delta v| +
\frac{1+\psi_v -\phi_v}{|\Delta v_{max}|}, \hfill
\end{equation}
where $\Delta v \in \left[0,\Delta v_{max} \right]$ with $\Delta v_{max}=v_p-v_*$ and the preferred velocity is given by
$v_p=H(\rho^a_{\theta_*}-\rho^a_\theta)$.
An explicit definition of $\Delta \theta_{max}$ and $\Delta v_{max}$ is reported in Fig.~\ref{fig:p_transizione}. 
In Eq.~\eqref{eq:C}, $\phi_\theta$ and $\psi_\theta$ give the negative and positive contributions to the trend to $v_p$,
respectively,
\begin{equation}
 \phi_v  =  \rho,  \qquad \hbox{and} \qquad \psi_v  =  \alpha u_* \Delta v_{max}.
\end{equation}
Finally let us consider the third step consisting in modeling how interactions modify the activity variable by increasing or decreasing it, 
but supposing that each pedestrian keeps one's own strategy without transition into another functional subsystem. 
The simplest assumption is that given a random initial condition of the distribution over such variable, then this is not modified by interactions. 
This assumption correspond to the following expression of the transition probability density
\begin{equation}
\cD(u_* \to u)= \delta_u.
\end{equation}

\subsection{Models in bounded domain}
When dealing with dynamics in a bounded domain, nonlocal interactions occur with walls, obstacles, and one or more exists.
The strategy pursued in our modeling approach is that the aforementioned interactions, which replace the classical local boundary
conditions, define a new preferred walking direction which is determined by a two-steps procedure. As a first step, the candidate pedestrian
changes in probability the direction of motion and 
the velocity modulus by following the same rules described in the preceding subsection, Eqs.~\eqref{eq:B} and \eqref{eq:C}, respectively.
As a second step, by keeping the same velocity modulus, the direction of motion is further changed so as to account for the presence of solid walls.
Accordingly, the probability transition density modifies to

\begin{equation}
\label{eq:transition_probability_bounded}
\cA_i[\f](t, \bx, \bw; \alpha) =
\cB^{(2)}(\theta^{(1)} \to \theta)
\cD [\rho] ( u_* \to u )
\cC [\rho] ( v_* \to v )
\cB_i^{(1)} [\rho,\mathbf{q}_i] (\theta_* \to \theta^{(1)}).
\end{equation}
where $\cB_i^{(1)}$ is defined by Eq.~\eqref{eq:B}, with $\theta^{(1)}$ in place of $\theta$, and $\cB^{(2)}$ reads

\begin{equation}
\label{eq:B_2}
 \cB^{(2)}(\theta^{(1)} \to \theta) = \delta (\theta-\theta^{(2)}(\theta^{(1)}))
\end{equation}
In Eq.~\eqref{eq:B_2}, the angle of motion $\theta^{(2)}$ gives the direction of the velocity when the latter is rotated so as to reduce its
normal component linearly with the distance from the wall.
More specifically, after the first step, the velocity components normal and tangential to the wall are given, respectively, by

\begin{eqnarray}
 v_n^{(1)} & = & (\bn \otimes \bn) \bv^{(1)} \\
 v_t^{(1)} & = & (\bI-\bn \otimes \bn) \bv^{(1)}
\end{eqnarray}
where $\bv^{(1)}$ is the velocity of the pedestrian after the decisional process that weigths target, stream and vacuum effects,
and $(\bn \otimes \bn), (\bI -\bn \otimes \bn)$ are the projection operators in the directions normal and tangential to the wall.
Due to the interaction with the wall, in a second step, the pedestrian velocity components are modified as follows

\begin{eqnarray}
\label{eq:vel_n}
 v_n^{(2)} & = & d_w v_n^{(1)} \\
\label{eq:vel_t}
 v_t^{(2)} & = & \mbox{sign}{(v_{t}^{(1)})}\left[ {v^{(1)}}^2 - {v_n^{(2)}}^{2} \right]^{1/2}
\end{eqnarray}
Eq.~\eqref{eq:vel_n} shows the the velocity component normal to the wall decreases linearly approching the wall and becomes nought
in the limiting case of a pedestrian in contact with the wall.
The tangential component, as given by Eq.~\eqref{eq:vel_t}, is obtained by requiring that the modulus of the velocity is constant.
The angle $\theta^{(2)}$ which enters in Eq.~\eqref{eq:B_2}, is thus given by

\begin{equation}
 \bv^{(2)} = v_n^{(2)} \bn + v_t^{(2)} (\bI-\bn \otimes \bn) = v^{(2)} (\cos{\theta^{(2)}}, \sin{\theta^{(2)}} )
\end{equation}

\subsection{Models in panic conditions}
The mathematical model proposed in the preceding subsections is such that pedestrians equally share the different trends,
however already paper~\cite{[ACK14]} observed that the presence of panic increases their attraction toward the stream effect.
Namely, pedestrians try excessively to do what the others do and  neglect the search of less congested areas.
This feature can be modeled by the parameter $\beta$ in Eq.~\eqref{nu}.
However, the attraction toward the stream is not the only effect of panic, which increases the walking ability up to a certain extent.
Summarizing, increasing the level of panic indices the following:

\begin{itemize}

\item $\beta \, \uparrow:$ Attraction to what the others do against the search of less crowded areas.

\vskip.2cm \item $u \, \uparrow:$ Increasing of the ability to express the walking ability.

\end{itemize}

If the crowd is in a sufficiently small domain, one can suppose that at a certain critical time $t_c$ the aforesaid effects are homogeneously captured by the whole population. Therefore a simplest approach consists in introducing a parameter $p$ suitable to define the level of panic, or other anomalous behaviors, by
$p \in [0,1]$, where $p=0$ corresponds to normal conditions, while $p=1$ to the highest admissible level. Then, the transition of the parameter of the model can be defined as follows:
\begin{eqnarray}
&& t \leq t_c\, : \, \beta_p = \beta, \quad u_p = u, \nonumber\\
&&\\
&& t > t_c\, : \, \beta_p = \beta + p(1 - \beta), \quad u_p = u[1 + p(1 - u)].\nonumber
\end{eqnarray}

On the other hand, in case of overcrowding in a large environment, one can figure out situations where panic, or any anomalous behavior, is localized in a small area and is transported in the whole domain. For instance:
$$
p>0, \quad \hbox{for} \quad \bx \in \Sigma_p \subset \Sigma; \quad p=0, \quad \hbox{for} \quad \bx \not\in \Sigma_p.
$$

The modeling approach can be developed by adding $p$ to the set, which defines the micro-state so that the distribution functions are $f_i=f_i(t,\,\bx,\,v, \theta, u,p)$, while the transition probability density is factorized as follows:
\begin{equation}
\label{eq:transition_probability2}
\cA [f] (\bw_* \to \bw) =   \cC [\rho] ( v_* \to v ) \cB [\rho] (\theta_* \to \theta)\cE [\rho] ( p_* \to p ) \cD [\rho] ( u_* \to u ),
\end{equation}
where now $\bw = \{v, \theta, u, p\}$. The modeling of $\cE$ can be based on a consensus and learning type  games, see Section 3 of \cite{[BKS13]}, while
the following adjustment of Eq.~(4.21)$_2$ can be used again

\section{Simulations}
This section develops some computation and simulations to test the capability of the model to depict emerging behaviors which are observed in reality
and are of interest in the evacuation dynamics. 
More specifically, we consider the problem of two groups of people walking in opposite directions in a crowded street~\cite{[HM95]}.
The analysis focuses in particular on the role of the parameter $\beta$ which is descriptive of the presence of panic conditions.
It is well understood that this case study do not cover the whole varieties of dynamics to be considered. 
However, it leads to a number of dynamical behaviors worth to be studied.
In Subsection~\ref{sub:particle_method}, we give a a brief overview of the numerical method adopted in the present study to solve Eq.~\eqref{eq:kin}.
The simulations results are then presented and discussed in Subsection~\ref{sub:numerical_results}.

\subsection{Outline of the numerical method}
\label{sub:particle_method}
Obtaining numerical solutions of Eq.~\eqref{eq:kin} is a challenging task because the unknown function depends, in principle, on seven variables.
Moreover, the computation of the interaction term on the right hand side requires the approximate evaluation of a multidimensional integral. 
Numerical methods for solving similar kinetic equations in rarefied gas dynamics studies can be roughly divided into three groups:
\begin{itemize}
 \item[(a)] Particle methods
 \item[(b)] Semi-regular methods
 \item[(c)] Regular methods  
\end{itemize}
Particle methods are by far the most popular and widely used simulation methods~\cite{[B94],[RW05]}. 
The basic idea is to represent the distribution function by a number of mathematical particles which move in the computational domain
and interact according to stochastic rules derived from the kinetic equation.
Macroscopic flow properties are obtained by means of weighted averages of the particle properties.
Methods in groups (b) and (c) adopt similar strategies in discretizing the distribution function on a regular grid in the phase space and in using 
finite difference schemes to approximate the streaming term on the l.h.s of Eq.~\eqref{eq:kin}. 
However, they differ in the way the interaction integral is evaluated.
In semi-regular methods, the interaction integral is computed by Monte Carlo or quasi Monte Carlo quadrature methods \cite{[F91]} 
whereas deterministic integration schemes are used in regular methods \cite{[A01],[GG14]}. \\
Kinetic equations for pedestrian dynamics are usually solved by means of regular methods of solution~\cite{[BBK13],[ACK14]}.
In the present work, however, Eq.~\eqref{eq:kin} is solved by using a Nanbu-like particle scheme.
Compared with different methods of solutions, the proposed approach provides some advantages such as the possibilities to easily deal with 
complex geometries as well as to account for a sophisticated behavioral decisional process. 

\begin{figure}[t!]
   \begin{center}
     \epsfig{file=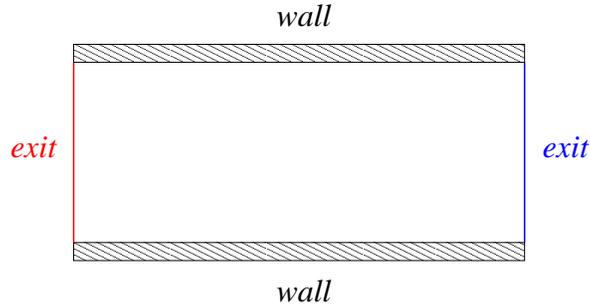,height=4cm} 
   \end{center}
\caption{Geometry of the case study.}
\label{fig:config}
\end{figure}

\subsection{Numerical results}
\label{sub:numerical_results}
In the following, simulations are carried out for two flows of pedestrians moving in opposite directions in a narrow 
street. The geometry of the problem is displayed in Fig.~\ref{fig:config}.
Initially, the two groups of pedestrians are uniformly distributed in a rectangular room of dimension $L_x\times L_y =20\,m\times4\,m$ 
which is open on the left and on the right sides. Periodic boundary conditions are assumed in the transversal direction. \\
The objectives of the simulations are the following:

\begin{enumerate}
 \item Describing the segregation of pedestrians into lanes of uniform walking direction and assess the influence of the 
       parameter $\beta$ and of the density $\rho$ on the shape of the pattern.
 \item Study of the evacuation dynamics for the two crowds. 
\end{enumerate} 
We first show that the model predicts the spontaneous formation of parallel lanes. 
The transient from disorder to order can be quantitatively assess by computing the band index, $Y_{\mbox{\tiny B}}(t)$ 
which measures the segregation of opposite flow directions~\cite{[Y98]}. In the present work it may be generalized to
the expression

\begin{equation}
 \label{eq:banda}
 Y_{\mbox{\tiny B}}(t) = \frac{1}{L_x L_y} \int_{0}^{L_x} \int_{0}^{L_y} 
                         \frac{| \rho_1 (t,\bx)-\rho_2 (t,\bx) |}{\rho_1 (t,\bx)+\rho_2 (t,\bx)} d\bx                
\end{equation}

\begin{figure}[t!]
\begin{subfigure}{.4\textwidth}
  \centering
  \includegraphics[width=1.0\linewidth]{band_rho10.eps}
\end{subfigure}%
\begin{subfigure}{.6\textwidth}
  \centering
  \includegraphics[width=1.0\linewidth]{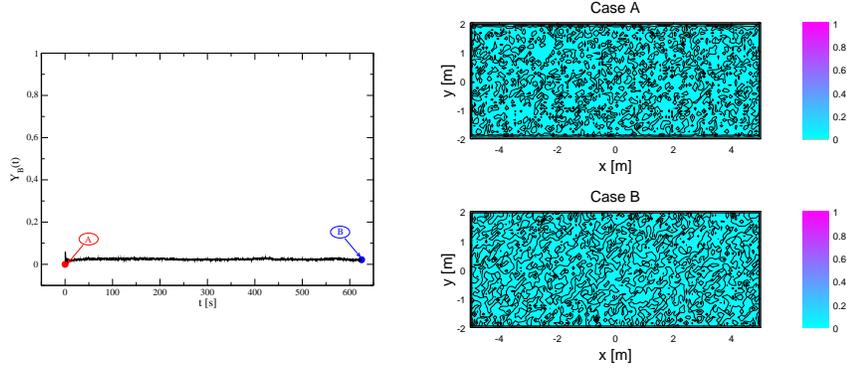}
\end{subfigure}
\caption{Bidirectional flow of $10$ pedestrians. 
         Left panel: Temporal evolution of the band index.
         Right panels: Contour plots of the pedestrian density at time $t=0 \, s$ (upper panel) and $t= 625 \, s$ (lower panel).}
\label{fig:segregation_rho10}
\end{figure}
\begin{figure}[t!]
\begin{subfigure}{.4\textwidth}
  \centering
  \includegraphics[width=1.0\linewidth]{band_rho150.eps}
\end{subfigure}%
\begin{subfigure}{.6\textwidth}
  \centering
  \includegraphics[width=1.0\linewidth]{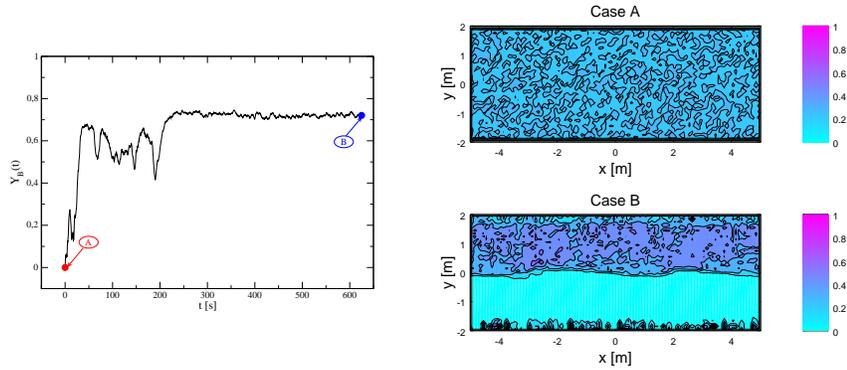}
\end{subfigure}
\caption{Bidirectional flow of $150$ pedestrians. 
         Left panel: Temporal evolution of the band index.
         Right panels: Contour plots of the pedestrian density at time $t=0 \, s$ (upper panel) and $t= 625 \, s$ (lower panel).}
\label{fig:segregation_rho150}
\end{figure}

\noindent
According to its definition, $Y_{\mbox{\tiny B}}(t)=0$ for mixed counterflows and $1$ for a perfect segregation of the opposite flows.
In a low density crowd of $10$ pedestrians, Fig.~\ref{fig:segregation_rho10}a shows that the band index is only slightly different from zero. 
Indeed pedestrians randomly fill the domain and no segregation takes place
as shown in Fig.~\ref{fig:segregation_rho10}b which reports snapshots of pedestrian density at time $t=0\, s$ (upper panel) and
$t=625\, s$ (lower panel). 
Figs.~\ref{fig:segregation_rho150}a and~\ref{fig:segregation_rho150}b show the same results but for a dense crowd of $150$ pedestrians.
In this case, the emergence of spatial segregation is apparent, namely the different groups of pedestrians form in line and file alternatively. 
It is worth pointing out that, unlike most of the previous studies on the subject, no specific repulsion forces have been assumed between
pedestrians belonging to different populations.
We then compute the ratio between the pedestrian mean flow rate through a section of the street and the number of pedestrian as a function
of the number of pedestrian for a different sensitivity to the stream with respect to the search of vacuum. 
Below a threshold value, $N_p \approx 125$, the reduced mean flow rate is almost constant for any value of $\beta$, whereas, as the number of
pedestrian increases, it reduces.
This is not unexpected since in the behavioral decisional process of pedestrians 
the weight of the target direction  decreases with 
the mean crowd density, $\bar{\rho}$. When $N_p \gtrsim 125$, which corresponds to $\bar{\rho} \gtrsim 1/2$, 
the stream and vacuum effects become dominant thus leading to a loss of efficiency in reaching the target.


\begin{figure}
 \centering
 \includegraphics[width=0.8\linewidth]{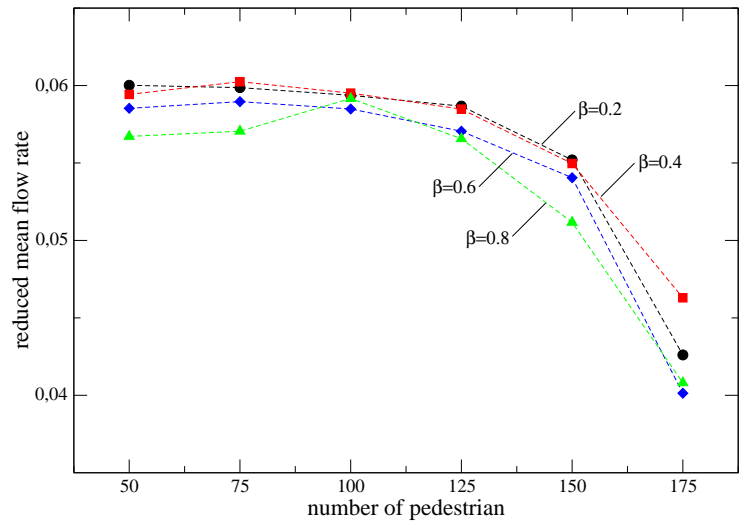}
 \caption{Ratio between the pedestrian mean flow rate through a section of the street and the number of pedestrian for different $\beta$.} 
\end{figure}

\newpage

\section{Some preliminary reasonings on further step towards a social crowd dynamics}

The sequential steps followed in this paper to pursue the objective of developing an approach to behavioral social dynamics can be listed as follows:

\begin{enumerate}

\item Definition of the features of behavioral dynamics;

\vskip.1cm \item Derivation of a general structure  suitable to offer  the conceptual basis for the derivation of models;

\vskip.1cm \item Modeling interactions at the microscopic scale to implement the aforesaid structure and derive specific models in unbounded and bounded domains;

\end{enumerate}

This process has been focused on a detailed analysis of the influence of panic conditions on the overall dynamics, which included some aspects of behavioral dynamics by the activity variable assumed to be heterogeneously distributed over individuals. It has been shown how this distribution evolves in time and space due to interactions. However, the approach does not yet tackle a challenging problem, which consists in modeling the dynamics of a crowd subdivided into different functional subsystems behaving with different features and purposes. In particular, the dynamics should predict transitions from one subsystem to the other.

It is worth mentioning that this type of investigation is motivated by security problems. As an example, one can consider a crowd of individuals in a public demonstration to support or make opposition to political issue. These individuals can be subdivided into  a large group manifest correctly their position, while a small group are rioters. Their number can grow in time due to interactions which might persuade the other part of the crowd to join them. Similarly a small group of security forces might react to provocations out of the settle protocol, but their number might also grow due to interactions, which cause an excess of reactions. The modeling  can take advantage of the more general structure given in Eqs.~(3.7), (3.10), (3.11), as well as of the literature on social dynamics \cite{[BHT13],[DL14]}. These brief  perspective ideas guide a research program already in progress.



\begin{thebibliography}{99}

\bibitem{[ABE08]}
    \newblock Ajmone~Marsan G., Bellomo N., and Egidi M.,
    \newblock Towards a mathematical theory of complex socio-economical systems by functional subsystems representation,
    \newblock {\em Kinet. Relat. Models}, {\bf 1} (2008), 249--278.

\bibitem{[ACK14]}
    \newblock Agnelli J-P., Colasuonno F., and Knopoff D.,
    \newblock A kinetic theory approach to the dynamics of crowd evacuation from bounded domains,
    \newblock {\em Math. Models Methods Appl. Sci.}, \textbf{25(1)} (2015).

\bibitem{[AP12]}
    \newblock Albi G. and  Pareschi L.,
    \newblock Modeling of self-organized systems interacting with a few individuals: From microscopic to macroscopic dynamics,
    \newblock {\em Appl. Math. Letters},  \textbf{26} (2013), 397--401.

\bibitem{[A01]} 
    \newblock V. V. Aristov,
    \newblock {\em Direct Methods for Solving the Boltzmann Equation and Study of Nonequilibrium Flows},
    \newblock Springer-Verlag, New York, 2001.

\bibitem{[BCC08]}
    \newblock Ballerini M., Cabibbo N., Candelier R., Cavagna A., Cisbani E., Giardina I., Lecomte V., Orlandi A., Parisi G., Procaccini A., Viale M., and Zdravkovic V.,
    \newblock Interaction ruling animal collective behavior depends on topological rather than metric distance: evidence from a field study,
    \newblock \textit{Proc. Natl. Acad. Sci. USA},  \textbf{105(4)} (2008), 1232--1237.

\bibitem{[BBK13]}
    \newblock Bellomo N., Bellouquid A.,  and Knopoff D.,
    \newblock From the micro-scale to collective crowd dynamics,
    \newblock {\em SIAM Multiscale Model. Simul.},  \textbf{11(3)} (2013), 943--963.

\bibitem{[BB14]}
    \newblock Bellomo N. and Bellouquid A., 
    \newblock On multiscale models of pedestrian crowds -  from mesoscopic to macroscopic,
    \newblock {\em Comm. Math. Sciences}, (2014), to appear.

\bibitem{[BHT13]}
    \newblock Bellomo N., Herrero M.A., and Tosin A.,
    \newblock On the dynamics of social conflicts looking for the Black Swan,
    \newblock {\em Kinet. Relat. Models},  \textbf{6} (2013), 459--479.

\bibitem{[BKS13]}
    \newblock Bellomo N., Knopoff D., and Soler J.,
    \newblock On the difficult interplay between life, ``complexity'', and mathematical sciences
    \newblock {\em Math. Models Methods Appl. Sci.},  \textbf{23} (10) (2013), 1861--1913.

\bibitem{[BPT12]}
    \newblock  Bellomo N.,  Piccoli B.,  and Tosin A.,
    \newblock Modeling crowd dynamics from a complex system viewpoint,
    \newblock \emph{Math. Models Methods Appl. Sci.}, \textbf{22} (2012), paper n. 1230004.

\bibitem{[BS12]}
    \newblock Bellomo N. and Soler J.,
    \newblock  On the mathematical theory of the dynamics of swarms viewed as complex systems,
    \newblock \emph{Math. Models Methods Appl. Sci.}, \textbf{22} (2012),  paper n. 1140006.

\bibitem{[B94]} Bird G.A.,
    \newblock {\em Molecular Gas Dynamics and the Direct Simulation of Gas Flows},
    \newblock Oxford University Press, 1994.
    
\bibitem{[DH03]}
    \newblock Daamen W.  and Hoogedorn S.P.,
    \newblock Experimental research of pedestrian walking behavior,
    \newblock \emph{TRB Annual Meeting  CD-ROM}, (2006).

\bibitem{[DL14]}
    \newblock Dolfin M. and Lachowicz M.,
    \newblock Modeling altruism and selfishness  in  welfare dynamics: The role of nonlinear interactions,
    \newblock \emph{Math. Models Methods Appl. Sci.}, \textbf{24} (2014), 2469--2482.

\bibitem{[FM15]}
    \newblock S.~Faure and B.~Maury,
    \newblock Crowd motion from the granular standpoint,
    \newblock {\em Math.  Models Methods Appl. Sci.},   \textbf{25} (2015).

\bibitem{[F91]} 
    \newblock Frezzotti A., 
    \newblock Numerical study of the strong evaporation of a binary mixture, 
    \newblock \emph{Fluid Dynamics Research}, \textbf{8} (1991), 175-187.
 
\bibitem{[GG14]}
    \newblock Ghiroldi G. P. and Gibelli L.,
    \newblock A direct method for the Boltzmann equation based on a pseudo-spectral velocity space discretization,
    \newblock \emph{J. Comput. Phys.}, \textbf{258} (2014), 568--584. 
 
\bibitem{[HEL01]}
    \newblock  Helbing D.,
    \newblock Traffic and related self-driven many-particle systems,
    \newblock   \emph{Rev. Modern Phys.},  \textbf{73} (2001), 1067--1141.

\bibitem{[HFV00]}
    \newblock Helbing D., Farkas I.,  and Vicsek T.,
    \newblock Simulating dynamical feature of escape panic,
    \newblock {\em Nature},  \textbf{407} (2000), 487--490.

\bibitem{[HJ09]}
    \newblock Helbing D. and Johansson A.,
    \newblock  Pedestrian crowd and evacuation dynamics,
    \newblock \emph{Enciclopedia of Complexity and System Science},  Springer, (2009), 6476--6495.

\bibitem{[HJA07]}
    \newblock Helbing D., Johansson A.,  and Al-Abideen H.Z.,
    \newblock Dynamics of crowd disasters: An empirical study,
    \newblock \emph{Phys. Rev. E},  \textbf{75} (2007), paper no.~046109.

\bibitem{[HM95]}
    \newblock Helbing D., Moln\'ar P.,
    \newblock Social forse model for pedestrian dyanmics,
    \newblock \emph{Phys. Rev. E},  \textbf{51} (5) (1995), 4282--4286.

\bibitem{[MHG09]}
    \newblock Moussa\"id M., Helbing D.,  Garnier  S., Johansson A., Combe  M., and Theraulaz G.,
    \newblock Experimental study of the behavioural mechanisms underlying  self-organization in human crowds,
    \newblock \emph{Proc. Roy. Soc. B},  \textbf{276} (2009), 2755--2762.

\bibitem{[MT11]}
    \newblock Moussa\"id M. and Theraulaz G.,
    \newblock Comment les pi\'etons marchent dans la foule.
    \newblock {\em La Recherche},   \textbf{450}, (2011), 56--59.

\bibitem{[NOW06]} Nowak M.A., 	
    \newblock {\em Evolutionary Dynamics: Exploring the Equations of Life},
    \newblock Belknap Press of Harvard University Press, (2006).

\bibitem{[RW05]} S. Rjasanow, W. Wagner,
    \newblock {\em Stochastic numerics for the Boltzmann equation},
    \newblock Springer, 2005.

\bibitem{[SS11]}
    \newblock Schadschneider A.,  and Seyfried A.,
    \newblock Empirical results for pedestrian dynamics and their implications for modeling,
    \newblock {\em Netw. Heterog. Media}, \textbf{6} (2011), 545--560.

\bibitem{[SSKB06]}
    \newblock Seyfried A., Steffen B., Klingsch W.,  and Boltes M.,
    \newblock The fundamental diagram of pedestrian movement revisited,
    \newblock {\em J. Stat. Mech.: Theory and Experiments}, \textbf{360} (2006), 232--238.

\bibitem{[Y98]}
    \newblock Yamori Y.,
    \newblock Going with the flow: Micro-macro dynamics in the macrobehavioral patterns of pedestrian crowds,
    \newblock  {\em Psychol. Rev.}, \textbf{105} (1998), 530--557.   
               
\end{thebibliography}
\end{document}